\documentclass[aps,twocolumn,showpacs]{revtex4}
\usepackage{graphicx,amsmath}

\newcommand{\eq}[1]{Eq.~(\ref{#1})}
\newcommand{\fig}[1]{Fig.~\ref{#1}}

\newcommand{\olcite}[1]{Ref.~\onlinecite{#1}}
\newcommand{\olcites}[1]{Refs.~\onlinecite{#1}}
\newcommand{\sect}[1]{Section~\ref{#1}}

\newcommand{\ver}{ {\bf r} }
\newcommand{\vom}{\hat{\omega}}
\newcommand{\reff}{c }
\newcommand{\kb}{k_{\rm B}}
\newcommand{\ms}{\mu^\star}
\newcommand{\pr}{P_L(\rho,\mu)}

\begin{document}

\title{First-order phase transitions in two-dimensional off-lattice liquid 
crystals}

\author{H. H. Wensink and R. L. C. Vink}
\affiliation{Institut f\"ur Theoretische Physik II: Weiche Materie, 
Heinrich-Heine-Universit\"at D\"usseldorf, Universit\"atsstra{\ss}e 1, 
40225
D\"usseldorf, Germany}

\date{\today}

\begin{abstract} We consider an {\it off-lattice} liquid crystal pair potential 
in strictly two dimensions. The potential is purely repulsive and short-ranged. 
Nevertheless, by means of a single parameter in the potential, the system is 
shown to undergo a first-order phase transition. The transition is studied using 
mean-field density functional theory, and shown to be of the 
isotropic-to-nematic kind. In addition, the theory predicts a large density gap 
between the two coexisting phases. The first-order nature of the transition is 
confirmed using computer simulation and finite-size scaling. Also presented is 
an analysis of the interface between the coexisting domains, including estimates 
of the line tension, as well as an investigation of anchoring effects. 
\end{abstract}

\pacs{64.70.Md, 02.70.-c, 61.30.Cz, 61.30.Hn }

\maketitle

\section{Introduction}

The phase behavior of liquid crystals in two dimensions continues to be an 
interesting topic. On the one hand, at least for {\it lattice} liquid crystals, 
there is a clear resemblance to the planar spin or XY-model 
\cite{kosterlitz:1974}. In fact, the Lebwohl-Lasher (LL) model 
\cite{physreva.6.426}, which is one of the standard liquid crystal models, maps 
exactly onto the XY-model in two dimensions. The XY-model does not support 
long-range order \cite{physrevlett.17.1133} and, consequently, long-range 
nematic order is believed to be absent in many two-dimensional (2D) liquid 
crystals as well \cite{physreva.31.1776, bates.frenkel:2000, 
lagomarsino.dogterom.ea:2003} (an exception being \olcite{physreva.27.1221}). In 
addition, the XY-model features a Kosterlitz-Thouless (KT) transition 
\cite{kosterlitz.thouless:1972, kosterlitz.thouless:1973}. Consequently, phase 
transitions in two-dimensional liquid crystals are often interpreted in terms of 
the KT scenario \cite{farinas-sanchez:031711, berche.paredes:2005, 
lagomarsino.dogterom.ea:2003, physreva.31.1776, bates.frenkel:2000}. In 
particular, the KT transition is a {\it continuous} transition, as opposed to 
first-order. As a result, the possibility of a first-order transition occurring 
in a two-dimensional liquid crystal, has received relatively little attention.
 
Interestingly, computer simulations of appropriately {\it generalized} XY-models 
have shown that the possibility of also a first-order transition occurring in 
these systems should be taken seriously \cite{physrevlett.52.1535, 
physrevlett.88.047203, physrevlett.70.1327}. More recently, these findings were 
put on firm mathematical ground by van~Enter and Shlosman 
\cite{enter.romano.ea:2006, enter.shlosman:2005, physrevlett.89.285702}. In 
particular, it was demonstrated that Hamiltonians of the form
\begin{equation}
\label{eq:xy}
 {\cal H}_{\rm gXY} = - \sum_{<i,j>} \left( 
 \frac{1 + \vom_i \cdot \vom_j}{2} \right)^p,
\end{equation}
undergo first-order phase transitions when $p$ is large 
\cite{physrevlett.89.285702}. In \eq{eq:xy}, the sum is over nearest neighbors 
on a square lattice, and $\vom_i$ are two-dimensional unit vectors. The usual 
XY-model (up to a trivial constant and multiplicative factor) is recovered for 
$p=1$; the generalization is to also consider $p \gg 1$. In computer 
simulations, the first-order transition was already observed at $p=50$ 
\cite{physrevlett.52.1535}. Of course, for small $p$, the KT-scenario is 
ultimately recovered again, for which the transition is continuous.

The results obtained for generalized XY-models have similar consequences 
for two-dimensional liquid crystals. This was recently demonstrated in 
\olcite{vink:2006*b} using a generalized version of the LL-model
\begin{equation}
\label{eq:ll}
 {\cal H}_{gLL} = - \sum_{<i,j>} | \vom_i \cdot \vom_j |^p, 
\end{equation} 
the essential difference with \eq{eq:xy} being inversion symmetry under $\vom_i 
\leftrightarrow - \vom_i$ of the particle orientations. In particular, it was 
shown that \eq{eq:ll} undergoes a first-order temperature-driven transition, 
from an isotropic to a quasi-nematic phase, provided $p$ is sufficiently large. 
Again, the threshold value is at $p \approx 50$ \cite{vink:2006*b}. In the 
isotropic phase, the orientational correlations decay exponentially; in the 
quasi-nematic phase, they decay algebraically. Both phases thus lack long-range 
order in the thermodynamic limit, in agreement with the Mermin-Wagner theorem 
\cite{physrevlett.17.1133}. Consequently, the nematic order parameter cannot be 
used to describe the first-order transition in \eq{eq:ll}. Instead, a valid 
order parameter is the energy density, which shows a ``jump'' at the transition 
temperature. By keeping the energy density fixed at some value in the 
coexistence region, phase coexistence between isotropic and quasi-nematic 
domains can be realized \cite{vink:2006*b}.

The aim of this paper is to investigate how robust these findings are when 
also {\it off-lattice} liquid crystals in two dimensions are considered. 
To this end, we formulate an {\it off-lattice} model of a liquid crystal, 
which is somewhat inspired by the lattice Hamiltonian of \eq{eq:ll}. The 
model will be presented in \sect{model}. Next, the phase behavior of this 
model is studied, using theory and simulation. Indeed, both the theory and 
the simulation find strong evidence for the existence of a first-order 
transition, including a pronounced coexistence region. The coexistence 
region will be analyzed in some detail, including estimates of the line 
tension between the coexisting domains. An important improvement over the 
lattice model of \eq{eq:ll} is that the transition in the {\it 
off-lattice} model is characterized by a ``jump'' in the particle density. 
In other words, phase coexistence can now be studied by keeping the 
overall particle density fixed at some appropriate value, rather than the 
energy density. This finding is relevant for possible experiments, where 
the condition of fixed density is rather easy to implement (in contrast, 
keeping the overall energy fixed in an experiment would be much more 
difficult). At the same time, we find that the analogue of the 
$p$-exponent in \eq{eq:ll} must be quite large, before the first-order 
transition begins to show-up. Whether such ``sharp'' interactions can be 
realized experimentally is not yet clear, but some suggestions are made 
toward the end of this paper.

\section{Off-lattice liquid crystal in two dimensions}
\label{model}

In this paper, we consider an ensemble of particles, whose positions and 
orientations are confined to a two-dimensional plane. The particles interact 
with each other via a pair potential $v_{ij}$ of the form
\begin{eqnarray}\label{eq:ol}
  v_{ij} &=& \epsilon \, \sigma_{ij} \, 
  \left( 1 - |\vom_i \cdot \vom_j|^p \right) \, 
  u(r), \\
  \sigma_{ij} &=& 1 + \nu \left[ (\vom_i \cdot \ver_{ij} / r)^2
  + (\vom_j \cdot \ver_{ij} / r)^2 \right], \label{sigma}
\end{eqnarray}
with $\ver_{ij} = \ver_j - \ver_i$, $r = |\ver_{ij}|$, $-1/2 < \nu <1/2$, and 
coupling parameter $\epsilon>0$. In what follows, factors of $\kb T$ are 
absorbed into the parameter $\epsilon$, with $T$ the temperature, and $\kb$ the 
Boltzmann constant. The quantity $\vom_i$ is a two-dimensional unit vector 
denoting the orientation of the $i$-th particle; $\ver_i$ is a two-dimensional 
vector denoting the coordinate of the center of mass of the $i$-th particle. The 
radial function $u(r)$ in \eq{eq:ol} is assumed to be strictly positive and 
short-ranged. Here, we take a simple step function
\begin{equation}
  u(r) =
  \begin{cases}
  1 & r<a, \\
  0 & {\rm otherwise},
  \end{cases} \label{step}
\end{equation}
with $a$ the particle diameter which will henceforth serve as our unit of 
length.

As for the lattice Hamiltonian of \eq{eq:ll}, the potential is constructed such 
that inversion symmetry is maintained. Note also that $v_{ij}$ is purely 
repulsive. Nevertheless, we expect a first-order phase transition to take place, 
either at low temperature, or at high density, provided $p$ is large. The 
purpose of the parameter $\nu$ is to introduce a coupling between the 
orientational and translational degrees of freedom. By setting $\nu=0$, no such 
coupling occurs, and the potential becomes separable. For this special case, 
Straley has rigorously proved the absence of long-range nematic order 
\cite{physreva.4.675}. As far as we know, the absence or presence of long-range 
order for the case $\nu \neq 0$ is still an open question. 

\section{Density functional theory}
\label{the}

Within density functional theory (DFT) the thermodynamics and structure of a 
fluid is described by a functional $\Omega[\rho]$ of the one-particle 
distribution $\rho(\ver, \vom)$. The density functional is such that it is 
minimized for a given ($\mu$, $A$, $T$) by the equilibrium one-particle 
distribution and the minimum value of the functional is the grand potential 
\cite{evans:1979} (in the present study in two dimensions, $A$ is the system 
area). Here we use a simple mean-field functional which, in the absence of an 
external potential, can be cast in the following form
\begin{eqnarray}
&&  \Omega[\rho] = \int d \ver d \vom \rho(\ver,\vom)  \left (  \ln
[ \mathcal{V}  \rho (\ver, \vom)] - 1 - \mu   \right ) \\ 
&& +  \frac{1}{2} \int
d \ver d \vom  \int d \ver ^{\prime} d \vom ^{\prime} v(\Delta \ver ;
\vom , \vom ^{\prime} ) \rho( \ver , \vom ) \rho( \ver ^{\prime}, \vom ^{\prime}), \nonumber \label{dft1}
\end{eqnarray}
with $\Omega$ expressed in units $\kb T$, $\Delta \ver = \ver^{\prime} - \ver$, 
$\mathcal{V}$ the 2D thermal volume of the particle and $\mu$ the chemical 
potential. The functional is known to give accurate results for dense fluids of 
soft spheres with bounded potentials \cite{grewe:1977, likos.ea:2001, 
archer:2005}. Recently, it was shown that the approach also works well for 
fluids of soft anisometric particles \cite{rex:2007}. The minimum condition on 
the functional, $\delta \Omega[\rho]/\delta \rho =0$, leads to a nonlinear 
integral equation
\begin{equation}
 \ln [ \mathcal{V}  \rho (\ver, \vom)] +   \int d \ver ^{\prime} d
 \vom ^{\prime} v(\Delta \ver ;
\vom , \vom^{\prime}) \rho( \ver ^{\prime}, \vom ^{\prime}) = \mu , \label{dft2}
\end{equation}
to be solved for the equilibrium distribution $\rho( \ver ^{\prime}, 
\vom^{\prime})$ at a given $\mu$.

\subsection{Bulk phase diagram}

Let us first focus on localizing the isotropic-to-nematic transition for bulk 
systems. As the average density in both states is spatially {\it uniform}, i.e. 
independent of $\ver$, we may write $\rho(\ver,\vom) = \rho f(\vom)$, where 
$\rho$ is the bulk density and $f$ an orientational distribution, subject to the 
normalization condition $\int d \vom f(\vom) =1$. In the isotropic (I) state, 
all orientations are equally probable and $f_{\text{I}}$ is a constant while in 
the nematic (N) phase $f_{\text{N}}$ is expected to be strongly peaked around 
the nematic director which we assume to be spatially uniform. This implies that 
long-range orientational order is always present in theory since fluctuations or 
local defects in the director field are not taken into account. Introducing the 
angle $\varphi \in [0,\pi]$ between the particle and the nematic director we may 
rewrite \eq{dft2} into a self-consistency equation for $f(\varphi)$
\begin{equation}
  f(\varphi) = Z^{-1}  \exp \left [- \rho \int d \varphi ^{\prime} E(\varphi ,
  \varphi ^{\prime})  f (\varphi ^{\prime}) \right  ]  \label{dft3},
\end{equation}
where $Z = \int d \varphi \exp[\cdots]$ to ensure normalization. Note that 
$f(\varphi)=f(\pi - \varphi)$ due to the inversion symmetry. The kernel 
$E(\varphi,\varphi^{\prime})$ is given by the following spatial integration
\begin{equation}
  E(\varphi,\varphi^{\prime}) = \int d \Delta \ver v(\Delta \ver ;
  \vom , \vom^{\prime}) \label{dft4}.
\end{equation}
The spatial integration in \eq{dft4} can then be carried out without difficulty 
to give
\begin{equation}
  E(\varphi,\varphi^{\prime}) = \epsilon (1+ \nu) \pi a^2 \left (  1 - | \cos
  (\varphi ^{\prime} - \varphi) |^p \right ). 
\end{equation}
With this result it is expedient to rewrite \eq{dft3} in the following way
\begin{equation}\label{dft5}
 f(\varphi) = Z^{-1}  \exp \left [ \reff \int d \varphi^{\prime} | \cos
 (\varphi^{\prime} - \varphi) |^p f (\varphi ^{\prime})   \right ], 
\end{equation}
where we have introduced the {\it effective} dimensionless density
\begin{equation}
 \reff = \pi(1+\nu)\rho a^2 \epsilon. \label{dfty}
\end{equation}
The combination $\rho \epsilon$ in \eq{dfty} illustrates the fact that phase 
transitions in our model can be brought about by either increasing the density 
or lowering the temperature. It is clear that the isotropic distribution $f_{I} 
= 1/\pi$ is a trivial solution of \eq{dft5} at all densities $\reff$. However at 
high densities non-trivial, i.e. nematic solutions are expected to branch off 
from the isotropic one. To locate the branching point (denoted by $\reff^\star$) 
we will perform a simple stability analysis of \eq{dft5}, along the lines of 
\olcites{kayser,mulder:1989}. Let us introduce the following expansions around the 
isotropic solution
\begin{eqnarray}
f(\varphi) &=& \frac{1}{\pi} \left [ 1 + \alpha a_1  \cos 2 \varphi +
  \alpha^2 a_2  \cos 4 \varphi  + \cdots \right ], \nonumber \\
\reff &=& \reff_0 + \alpha \reff_1  + \alpha^2 \reff_2 + \cdots \label{dft6},
\end{eqnarray}
in terms of a single order parameter $\alpha$. Likewise, we may expand the 
kernel as follows
\begin{equation}
 |\cos (\varphi^{\prime} - \varphi)|^p = \sum _{n \ge 0} k_{2n}(p)
 \cos (2n \varphi ) \cos (2n \varphi ^{\prime}), \label{kernel}
\end{equation}
with coefficients
\begin{eqnarray}\label{coef}
 k_{2n}(p) = \hspace{5cm} \\ 
 \frac{4}{\pi ^2} \int d \varphi \int d \varphi^{\prime}
 |\cos (\varphi^{\prime} - \varphi)|^p \cos (2n \varphi ) 
 \cos (2n \varphi^{\prime}). \nonumber
\end{eqnarray}
Inserting all expansions back into \eq{dft5} and keeping all contributions up to 
$\cal{O}(\alpha)$ gives the branching or bifurcation point
\begin{equation}
 \reff^\star = \reff_0 = 2/k_{2}(p) \label{dft7},
\end{equation}
implying that stable nematic solutions are expected for $\reff > \reff^\star$. 
We remark that for $p=1$ the branching density $c^\star = 3 \pi/2$ is identical 
to that of the 2D Onsager theory for infinitely thin hard needles 
\cite{physreva.4.675, note1}.

To verify the thermodynamic stability of the nematic solutions close to the 
branching density we must analyze its free energy. The dimensionless Helmholtz 
free energy (ignoring all irrelevant constants) is given by
\begin{equation}
 F[f]/N \sim \langle \ln [\reff \pi f(\varphi)] \rangle + \frac{\reff}{2} \langle 
 \langle (1 - |\cos (\varphi ^{\prime} - \varphi)|^p ) \rangle \rangle \label{dft8},
\end{equation}
where $\langle \cdots \rangle = \int d \varphi f(\varphi)$. The free energy 
difference $ \Delta F = F_{\text{N}}-F_{\text{I}}$ between the nematic and 
isotropic states close to the branching point can be written in Landau form in 
terms of the nematic order parameter $\alpha$
\begin{equation}
 \Delta F/N = A \alpha^2 + B \alpha^3 + C \alpha^4.
\end{equation} 
The coefficients can be obtained from extending the bifurcation analysis to 
higher order in $\alpha$ [using the expansions \eq{dft6} and \eq{kernel}] and 
performing an order-by-order solution of \eq{dft5} up to ${\cal O}(\alpha^{3})$. 
The algebra is straightforward but tedious and we will only give the final 
outcome here. It turns out that $A,B=0$ while
\begin{equation}
 C = -\frac{k_2 (p) - 2 k_{4} (p)}{64 [ k_2 (p) - k_4 (p) ]}. \label{dft9}
\end{equation} 
The order of the transition depends on the sign of $C$. If it is negative, small 
nematic perturbations around the branching point immediately stabilize the 
system and the transition is continuous. If $C$ is positive the incipient free 
energy difference goes up for small $\alpha$ which means that the actual phase 
transition must involve a density jump and is first-order. In the latter case 
the coexisting densities are indicated by binodal curves which can be computed 
in the usual way by requiring the pressure and chemical potential to be equal in 
the coexisting phases. The equilibrium $f(\varphi)$ for a given nematic density 
$\reff$ is obtained numerically from the consistency equation \eq{dft5} by 
dividing the interval $[0,\pi/2]$ into 100 equidistant grid points and employing 
the iteration scheme outlined in \olcite{herzfeld:1984}.

\begin{figure}
\begin{center}
\includegraphics[clip=,width=\columnwidth]{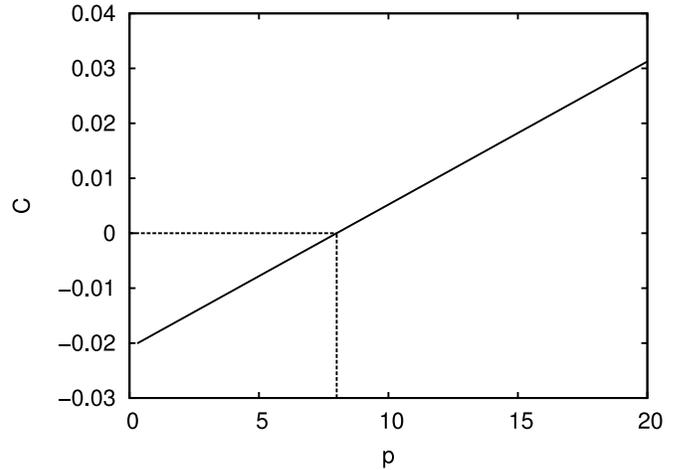}
\caption{\label{dft-df} Landau coefficient $C$ [\eq{dft9}] versus $p$. For $p>8$ 
the isotropic-nematic transition is first-order; for smaller $p$, it is 
continuous.}
\end{center}
\end{figure}

\begin{figure}
\begin{center}
\includegraphics[clip=,width=\columnwidth]{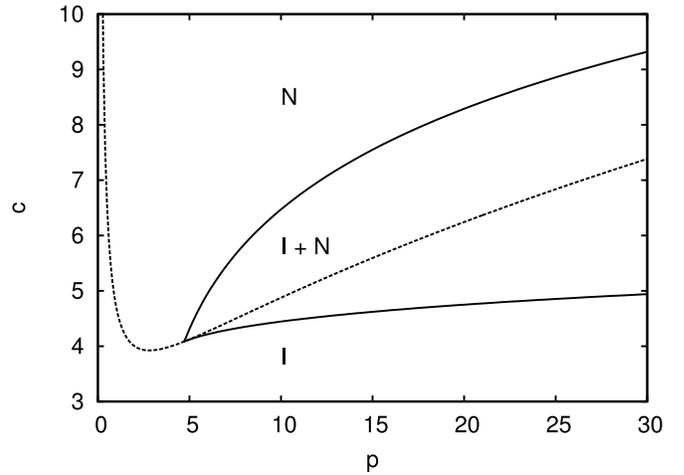}
\caption{\label{dft-phase} Theoretical phase diagram. The dashed curve 
represents the nematic branching line $\reff^\star$, calculated from \eq{dft7}.  
The binodals are given by the solid curves. A tri-critical point is located at 
$p=4.7$.}
\end{center}
\end{figure}

\begin{figure}
\begin{center}
\includegraphics[clip=,width=\columnwidth]{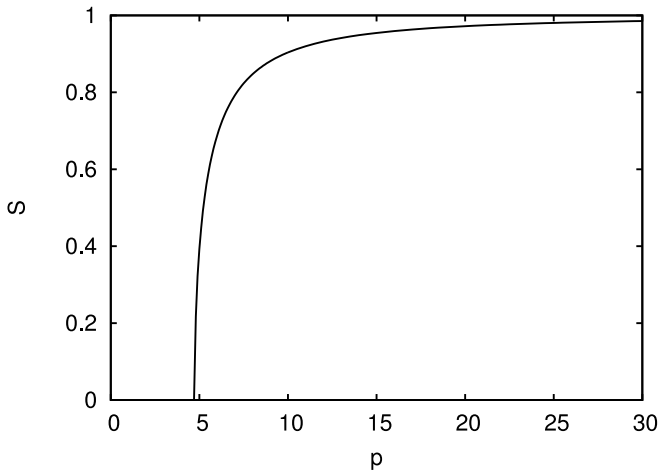}
\caption{\label{dft-s} Nematic order parameter $S = \langle \cos 2 \varphi 
\rangle $ corresponding to the nematic binodal in \fig{dft-phase}.}
\end{center}
\end{figure}

\fig{dft-df} shows that the isotropic-nematic transition is continuous for small 
$p$ but becomes first-order for $p>8$. The full phase diagram in \fig{dft-phase} 
shows that the actual crossover from continuous to first-order (marked by the 
tri-critical point where all phase lines meet) is located at a somewhat lower 
value for $p$, namely $p=4.7$. The discrepancy is expected since the bifurcation 
analysis usually provides an {\it upper} estimate for the critical points. We 
also remark that the non-monotonic behavior of the critical density at low $p$ 
is consistent with the simulation results of the generalized LL-model reported 
in \olcite{vink:2006*b}. The steep increase of the nematic order parameter in 
\fig{dft-s} suggests that a considerable degree of nematic order is expected in 
the coexisting nematic phase at large~$p$. 

\subsubsection{Asymptotic results for large $p$}

\begin{figure}
\begin{center}
\includegraphics[clip=,width=\columnwidth]{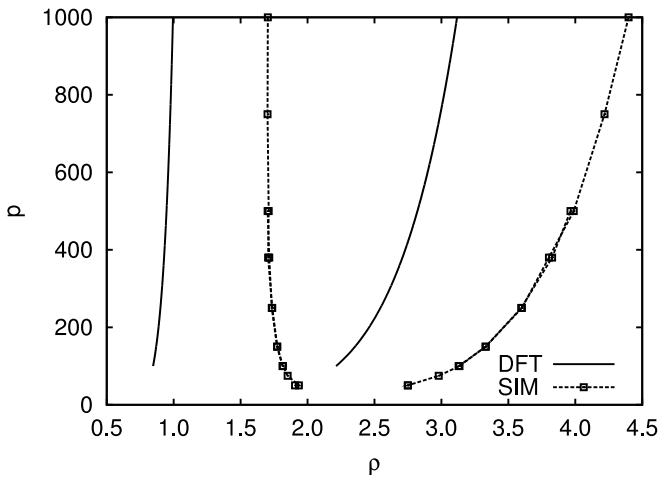}
\caption{\label{dft-gaussbino} Phase diagram of the two-dimensional liquid 
crystal model of \eq{eq:ol} using $\epsilon=2.5$. The solid curves show the 
theoretical binodals, obtained using the large $p$ approximation. Plotted is the 
dimensionless density $\rho=Na^2/A$ of the isotropic phase (left curve) and 
the nematic phase (right) versus $p$. Also shown are the corresponding 
simulation results (squares), where the dashed lines serve to guide the eye.}
\end{center}
\end{figure}

\begin{figure}
\begin{center}
\includegraphics[clip=,width=\columnwidth]{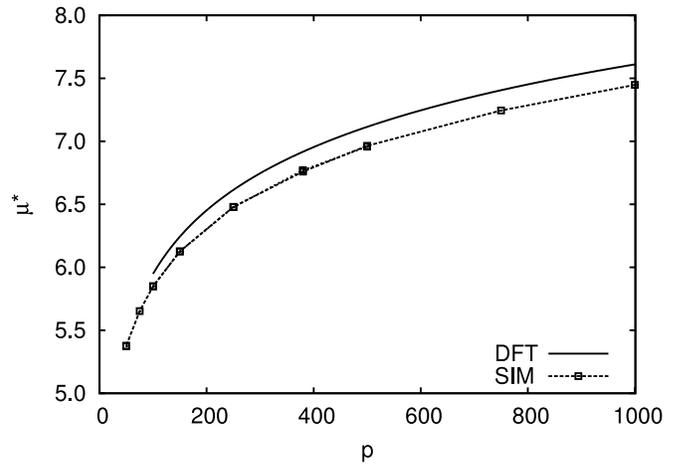}
\caption{\label{dft-gaussmu} Coexistence chemical potential $\ms$ versus $p$ 
corresponding to \fig{dft-gaussbino}. The solid curve is the theoretical 
result; squares are simulation results.}
\end{center}
\end{figure}

For very large $p$ (say larger than 100) the solution of \eq{dft5} on a grid 
becomes numerically awkward since $f(\varphi)$ gets extremely peaked at 
$\varphi=0$ (and $\pi$). It is therefore tempting to formulate a simple 
variational theory that allows us to access the phase diagram at asymptotically 
large $p$. Indeed, for highly nematic states $f(\varphi)$ is well-described by a 
gaussian trial function with variational parameter $\beta \gg 1$ 
\cite{odijk.hnwl:1985},
\begin{equation}
 f(\varphi) \sim \left ( \frac{2 \beta}{\pi}\right )^{1/2} \exp \left
 [- \frac{1}{2} \beta \varphi^2 \right ], \hspace{0.2cm} \text{for} 
 \hspace{0.2cm} 0 \le \varphi \le \frac{\pi}{2},
\end{equation}
and its mirrored version $f(\pi-\varphi)$ for the interval $\pi/2 \le \varphi 
\le \pi$. Inserting the gaussian into \eq{dft8} and integrating yields the 
following asymptotic result for the ideal free energy (first term)
\begin{eqnarray}
 \langle \ln [\reff \pi f(\varphi)] \rangle  &\sim & \ln \reff + \frac{1}{2}
 \ln 2 \pi \beta - \frac{1}{2}, \hspace{0.5cm} \text{(N)} \nonumber \\
 & = & \ln \reff, \hspace{2.9cm} \text{(I)}
\end{eqnarray}
valid for large $\beta$. Using the approximation
\begin{equation}
 |\cos x |^p \approx \exp [-(1/2) p x^2], \hspace{0.5cm} (p \gg 1) \label{approxker}
\end{equation}
in the excess free energy  (second term) allows us to calculate the double 
orientational averages  analytically
\begin{eqnarray}
 \langle \langle  |\cos (\varphi^{\prime} - \varphi)|^p  \rangle
 \rangle &\sim& \left ( \frac{\beta}{\beta + 2p} \right )^{1/2},
 \hspace{0.8cm} \text{(N)}
 \nonumber \\
 &\sim& \left ( \frac{2}{\pi p} \right )^{1/2}. \hspace{1.5cm} \text{(I)}
\end{eqnarray}
The value of $\beta$ is fixed (at a given density) by minimizing the
total nematic free energy with respect to the variational parameter. Some
rearranging then leads to the following minimization condition
\begin{equation}
  \left ( \frac{\beta}{p} + 2 \right )^3 - \reff^2  \frac{\beta}{p} = 0,
\end{equation}
which has to be solved numerically along with the coexistence equations for the 
chemical potential and pressure at a given $p$. These follow straightforwardly 
from the free energy \eq{dft8}. The resulting binodals and coexistence chemical 
potential are shown in \fig{dft-gaussbino} and \fig{dft-gaussmu}, respectively.

Another feature we want to point out is that the bifurcation density 
$\reff^\star$ at large $p$ scales as $\reff^\star \propto p^{1/2}$. The scaling 
relation can be easily established by making an asymptotic expansion of $k_2$ 
from \eq{coef}. For the regime $p \gg 1$ one can show with the aid of 
\eq{approxker} that
\begin{eqnarray}
  k_{2}(p) &\propto& \int _{0}^{\infty} d (\varphi^{\prime} - \varphi) \exp [
  -(1/2) p (\varphi^{\prime} - \varphi)^2 ], \nonumber \\ 
  &\propto& p^{-1/2},
\end{eqnarray}
up to leading order, hence $\reff^\star \propto p^{1/2}$. This result is 
analogous to the scaling of the critical coupling constant in the LL-simulations 
of \olcite{vink:2006*b}.

\subsection{Isotropic-nematic interface}

To assess the properties of the interface between the coexisting isotropic and 
nematic phases we have to go back to our initial DFT formulation in \sect{the}. 
If we assume the interface to be flat with a surface normal $\hat{x}$, the 
one-particle density will be non-uniform along this direction and depend on the 
spatial coordinate $x = \ver \cdot \hat{x}$ and angle $\varphi$. It is a 
solution of the integral equation \eq{dft2} with $\mu=\ms$, 
\begin{eqnarray}\label{dft10}
 \ln [ {\mathcal{V} } \rho (x, \varphi) ] + \hspace{5cm} \\
 \int d x ^{\prime} d \varphi ^{\prime} E_{x}( \Delta x ;
 \varphi , \varphi^{\prime} ; \vartheta ) \rho 
 ( x ^{\prime}, \varphi ^{\prime}) = \mu^{\star}, \nonumber 
\end{eqnarray}
$\ms$ being the chemical potential at coexistence, and $\Delta x = x^{\prime} - 
x$, subject to the boundary conditions
\begin{eqnarray}
 \rho (x, \varphi) &=& 
 \rho_{I}f_{I}, \hspace{1.2cm} (x \to -\infty), \nonumber \\
 \rho (x, \varphi) &=& 
 \rho_{N}f_{N}(\varphi), \hspace{0.5cm} (x \to \infty), \label{dft11}
\end{eqnarray}
(recall that $\varphi = \arccos (\hat{n} \cdot \hat{w})$). In \eq{dft10}, the 
kernel $E_{x}$ is defined as the pair potential (at fixed orientations) averaged 
over the distance $\Delta y$ perpendicular to $\hat{x}$ along which the system 
is homogeneous
\begin{equation}\label{dft12}
 E_{x}(\Delta x ; \varphi , \varphi^{\prime} ; \vartheta ) =
 \int_{-\infty}^{\infty} d \Delta y  v(\Delta \ver ;
 \varphi , \varphi^{\prime} ; \vartheta ).
\end{equation}
Note that because of the broken spatial symmetry the integral depends implicitly 
on the {\it anchoring angle} $\vartheta = \arccos (\hat{n} \cdot \hat{x})$ 
between the nematic director $\hat{n}$ and the surface normal. This becomes 
manifest when we focus on the translation-rotation coupling contribution 
$\sigma$ in \eq{sigma}. In explicit form it reads
\begin{equation}
 \sigma = 1 + \nu \left [ (  \Delta \hat { \ver } \cdot
 \mathcal{R}_{\vartheta} \vom )^2  + 
 ( \Delta \hat {\ver} \cdot \mathcal{R}_{\vartheta} \vom^{\prime} )^2 \right ],
\end{equation}
with $\Delta \hat {\ver} = \{\Delta x, \Delta y\}/(\Delta x ^2 + \Delta y 
^2)^{1/2}$ the center-of-mass difference unit vector, $\vom = \{ \cos \varphi, 
\sin \varphi \}$ and $\mathcal{R}_{\vartheta}$ the rotation matrix
\begin{equation}
 \mathcal{R}_{\vartheta} =
  \begin{pmatrix}
  \cos \vartheta & - \sin \vartheta \\ \sin \vartheta & \cos \vartheta
  \end{pmatrix}.
\end{equation}
Clearly, if $\nu=0$ there is no dependence on $\vartheta$ and the interfacial 
profiles are {\it identical} for all anchoring angles \cite{note2}. If $\nu \neq 
0$ the translational and rotational degrees of freedom are coupled and the 
interfacial properties will in general be dependent upon the anchoring angle. In 
particular we are interested in the line tension $\gamma$ which can be 
extracted from the equilibrium interfacial density profile. Inserting the 
one-body density into the functional \eq{dft1} yields the minimum grand 
potential
\begin{eqnarray}
 \Omega _{\rm min} = \hspace{5cm} \\
 \int d \ver \int  d \varphi \rho ( \ver,
 \varphi) \left \{  \frac{1}{2} \ln  [ \mathcal{V} \rho ( \ver,
 \varphi) ] -1 - \frac{1}{2} \mu^{\star} \right \}. \nonumber
\end{eqnarray}
The line tension $\gamma [ \mu ^{\star}(p),\nu, \vartheta ]$ is then 
obtained from the standard thermodynamic relation $\gamma = (\Omega_{\rm min} + 
PA)/L$, with $P$ the coexistence pressure.

\begin{figure}
\begin{center}
\includegraphics[clip=,width=\columnwidth]{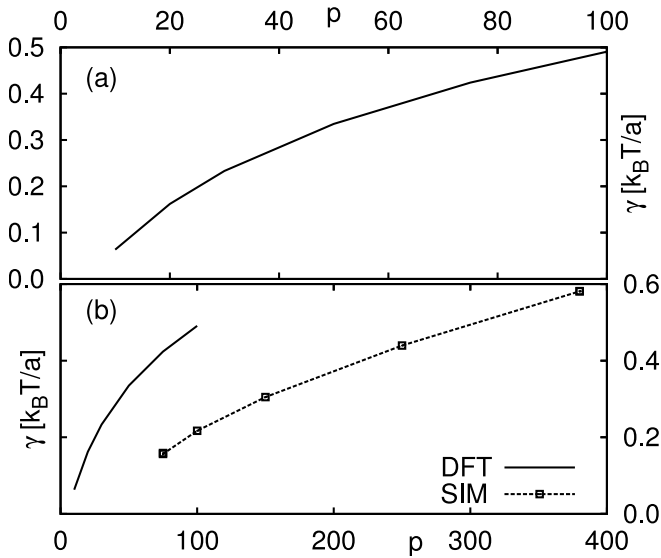}
\caption{\label{dft-gvsp} Line tension $\gamma$ (in units $\kb T/a$) for 
$\epsilon=2.5$ versus $p$. (a) DFT result. (b) Computer simulation results 
(squares), together with the DFT result (solid curve) for comparison. Note the 
different $p$-range in the above plots.}
\end{center}
\end{figure}

\begin{figure}
\begin{center}
\includegraphics[clip=,width=\columnwidth]{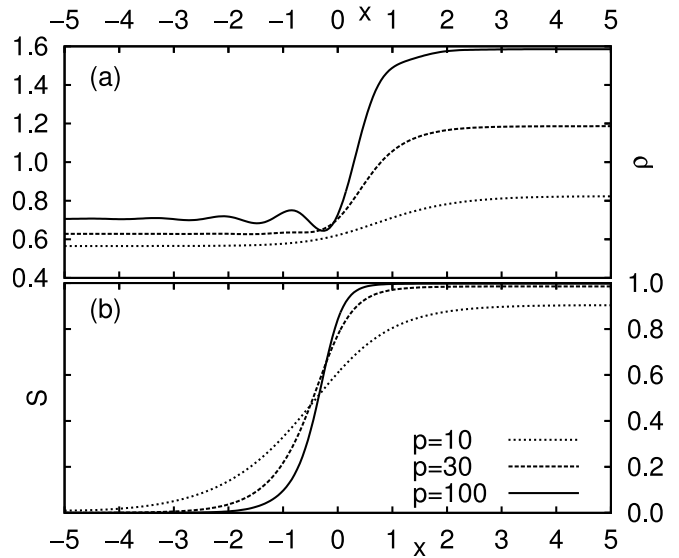}
\caption{\label{dft-profs} Interfacial profiles for (a) the density $\rho(x)$ 
and (b) the nematic order parameter $S(x)$ for $\nu = 0$. }
\end{center}
\end{figure}

In \fig{dft-gvsp} we show the line tension for $\nu=0$. To facilitate 
comparison with simulations later on we will henceforth fix the coupling 
parameter to $\epsilon=2.5$. Note that, owing to \eq{dfty}, changing this value 
does not give qualitatively different results but merely constitutes a linear 
shift in $\gamma$ and $\rho$. The increase of $\gamma$ as a function of $p$ 
reflects the transition becoming strongly first-order at large $p$. The 
corresponding interfacial profiles for the density $\rho(x)$ and the nematic 
order parameter $S(x)$, defined as
\begin{eqnarray}
 \rho(x) &=& \int d \varphi \rho(x,\varphi), \nonumber \\
 S(x)  &=& \rho ^{-1}(x) \int d \varphi \rho(x,\varphi) \cos 2 \varphi,
\end{eqnarray}
are shown in \fig{dft-profs}. As expected, the interface becomes sharper for 
large $p$. For $p=100$ small density oscillations on the isotropic side occur,
which point to a layering effect induced by the interface.

\begin{figure}
\begin{center}
\includegraphics[clip=,width=\columnwidth]{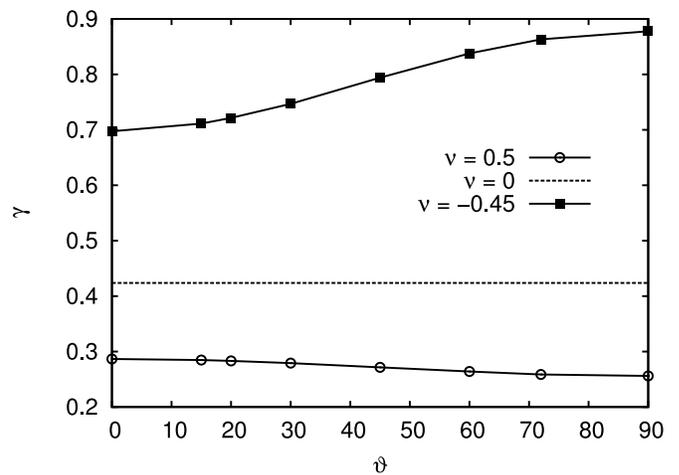}
\caption{\label{dft-anchor} Line tension $\gamma$ (in units $\kb T/a$) versus 
anchoring angle $\vartheta$ (in degrees), for several values of $\nu$ as 
indicated.}
\end{center}
\end{figure}

\begin{figure}
\begin{center}
\includegraphics[clip=,width=\columnwidth]{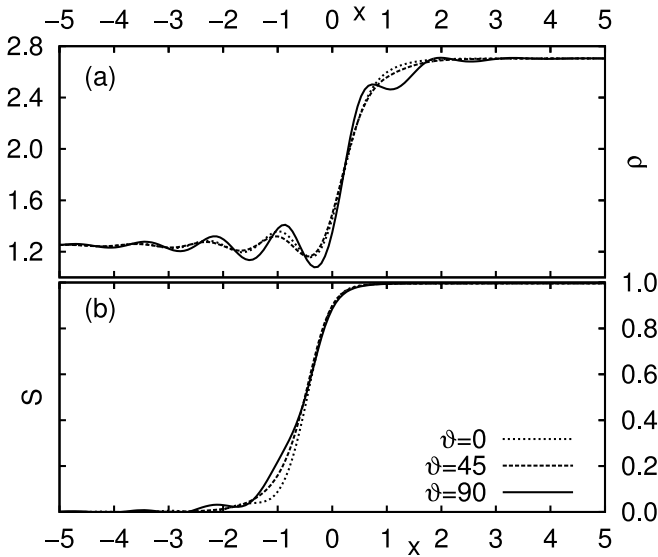}
\caption{\label{dft-aprofs} Interfacial profiles for $p=75$, $\nu = -0.45$ at 
different anchoring angles $\vartheta$ (in degrees).}
\end{center}
\end{figure}

Let us now focus on the anchoring behavior for a fixed value of $p=75$ and $\nu 
\neq 0$. The results, summarized in \fig{dft-anchor}, show a strong dependency 
of the line tension on the anchoring angle $\vartheta$, especially for $\nu < 
0$. Recall that end-to-end pair configurations are energetically more favorable 
than side-to-side ones in this case. While for $\nu<0$ the minimal tension is at 
$\vartheta=0$ implying {\it perpendicular} surface anchoring, a positive $\nu$ 
seems to be associated with {\it parallel} anchoring. Investigations for other 
$p$ reveal that this phenomenon is robust. Microscopic information extracted 
from the profiles for $\nu=-0.45$ is shown in \fig{dft-aprofs}. These reveal 
that the layering effect is considerably influenced by the anchoring angle. In 
particular, particles show enhanced localization across the interface if the 
nematic director is forced to be parallel to the interface.

\begin{figure}
\begin{center}
\includegraphics[clip=,width=\columnwidth]{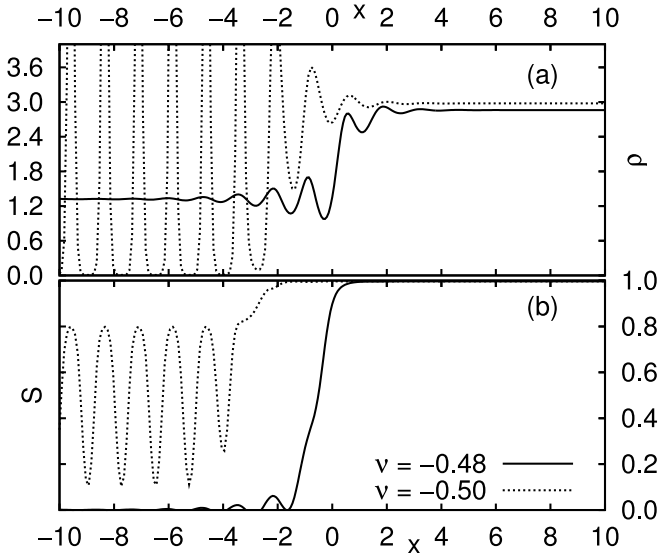}
\caption{\label{dft-nuprofs} Interfacial profiles for $p=75$ and anchoring angle 
$\vartheta = 90$ degrees.}
\end{center}
\end{figure}

An even more dramatic effect is encountered if we lower $\nu \to -0.5$, see 
\fig{dft-nuprofs}. Note that, for $\nu = -0.5$, the end-to-end configurations 
have zero repulsion and are therefore strongly favored. This explains the 
tendency of the system to form string like clusters along the interface as 
indicated by the sharp density peaks in \fig{dft-nuprofs}. However $\nu=-0.5$ 
seems to be a rather pathological case. The density modulations penetrate deeply 
into the isotropic bulk which raises strong suspicions as to whether the 
isotropic fluid state is stable against clustering or crystallization. It 
remains to be checked by simulations whether the IN transition is really 
pre-empted by a freezing transition in this case.

\section{Computer Simulation}

Next, we will confirm some of the theoretical results using computer simulation. 
To this end, grand canonical (GC) Monte Carlo simulations 
\cite{frenkel.smit:2001, landau.binder:2000} are performed, at chemical 
potential $\mu$, for the model of \eq{eq:ol} using $\epsilon=2.5$. Again, for 
the radial part, we take the step function of \eq{step}. In this work, we 
restrict the simulations to the case $\nu=0$ in \eq{eq:ol}. A computer 
simulation study of anchoring effects, predicted by the theory to occur when 
$\nu \neq 0$, will be postponed to a future publication.

\subsection{Finite-size scaling results for $p=75$}

We begin the simulations using $p=75$ in \eq{eq:ol}. Inspired by our theoretical 
results, we expect that, for $p=75$, a first-order phase transition should 
occur, when the chemical potential is tuned to its coexistence value $\ms$. In 
GC simulations, $\ms$ is determined from the distribution $\pr$, defined as the 
probability to observe a particle density $\rho$ in the system. In order to also 
simulate the regions of low probability, a biased sampling scheme is implemented 
\cite{virnau.muller:2004, vink.wolfsheimer.ea:2005}. The distribution $\pr$ 
depends on the chemical potential $\mu$, as well as on the size of the $L \times 
L$ simulation square (periodic boundary conditions are assumed). At $\mu = \ms$, 
$\pr$ becomes bimodal, with two peaks of equal area. In computer simulations, 
the transition can thus be located by varying $\mu$ until the ``equal-area'' 
criterion is obeyed.

\begin{figure}
\begin{center}
\includegraphics[clip=,width=\columnwidth]{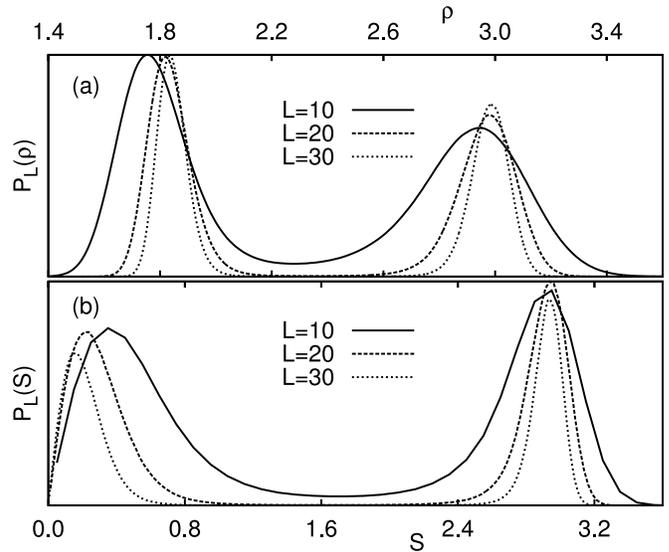}
\caption{\label{pn}(a) Coexistence distributions $\pr$ for various 
system sizes $L$ as indicated, with $\rho = Na^2/A$ the dimensionless particle 
density. The coexistence between the two phases is manifested by ``equal-area'' 
under the peaks. (b) The corresponding distributions $P_L(S)$ of the nematic 
order parameter $S$, given by \eq{eq:qq}.}
\end{center}
\end{figure}

\begin{figure}
\begin{center}
\includegraphics[clip=,width=\columnwidth]{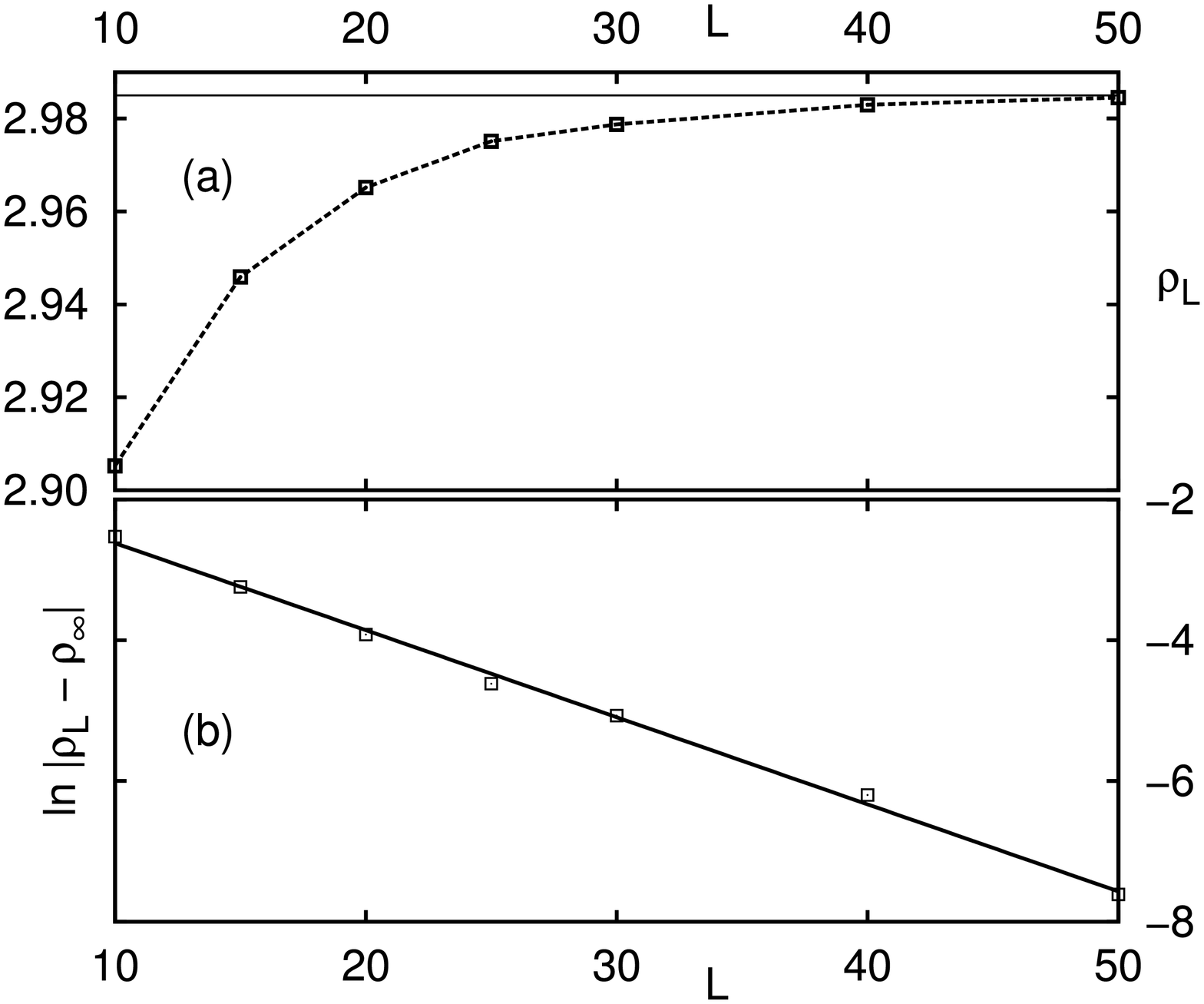}
\caption{\label{nemrho} Finite-size scaling of the position of the high-density 
peak in $\pr$ of \fig{pn}(a). (a) Peak position $\rho_L$ as a function 
of $L$. The horizontal line marks the density $\rho_\infty$ in the thermodynamic 
limit. (b) Finite-size scaling analysis using \eq{eq:sc}. The straight line is a 
linear fit; see details in text.}
\end{center}
\end{figure}

\begin{figure}
\begin{center}
\includegraphics[clip=,width=\columnwidth]{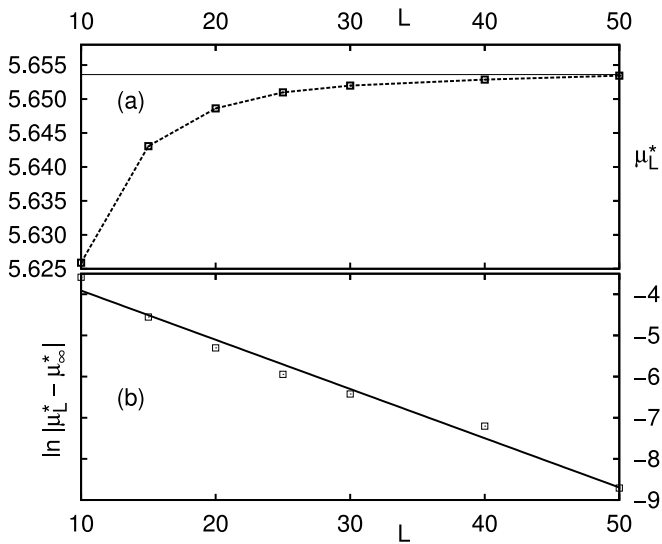}
\caption{\label{fss_mhu} Finite-size scaling of the coexistence chemical 
potential. (a) $\ms_L$ as a function of $L$. The horizontal line marks the 
coexistence chemical potential $\mu^*_\infty$ in the thermodynamic limit. (b) 
Finite-size scaling analysis using \eq{eq:sc}. The straight line is a linear 
fit; see details in text.}
\end{center}
\end{figure}

Indeed, we find that bimodal density distributions can be realized in this way. 
\fig{pn}(a) gives some examples for various system sizes $L$. By increasing $L$, 
the peaks become more narrow. This is to be expected because, in the 
thermodynamic limit $L \to \infty$, one has a sharp transition, and, 
consequently, a distribution featuring two $\delta$-peaks. In addition, closer 
inspection of \fig{pn}(a) also reveals a finite-size effect in the peak {\it 
positions}. Finite-size effects at first-order phase transitions have received 
considerable attention \cite{vollmayr.reger.ea:1993, binder.landau:1984, 
borgs.kotecky:1990}. We believe that the current state-of-the-art in this field 
is the rigorous treatment of Borgs and Kotecky \cite{borgs.kotecky:1990}. More 
precisely, for two-phase coexistence data obtained using the ``equal-area'' 
rule, an exponential $L$-dependence is predicted
\begin{equation}
\label{eq:sc}
 R \equiv |X_L - X_\infty| \leq {\cal O}( e^{-\tau L} ).
\end{equation}
Here, $X_L$ is the property of interest as obtained in the finite system of size 
$L$, $X_\infty$ the corresponding value in the thermodynamic limit, and $\tau$ a 
constant.

In \fig{nemrho}(a), we show the position $\rho_L$ of the high-density (nematic) 
peak of $\pr$ as a function of $L$. Next, we use \eq{eq:sc} to 
estimate the density $\rho_\infty$ of the nematic phase in the thermodynamic 
limit. To this end we have plotted, in \fig{nemrho}(b), the logarithm of $R$ as 
a function of system size $L$. Here, $\rho_\infty$ was taken to be a fit 
parameter, and tuned until the graph of $\ln R$ versus $L$ became linear. For 
$\rho_\infty \approx 2.985$, we find that $\ln R$ indeed becomes linear in $L$, 
which thus serves as our best estimate of the nematic density in the 
thermodynamic limit. For completeness, this estimate has also been marked in 
\fig{nemrho}(a) as the horizontal line. The density of the low-density 
(isotropic) peak can be obtained analogously. The corresponding scaling plot is 
qualitatively similar to \fig{nemrho} and therefore not shown here. For the 
density of the isotropic phase in the thermodynamic limit, we obtain 
$\rho_\infty \approx 1.850$.

In addition, we also observe an $L$-dependence in the coexistence chemical 
potential $\ms_L$. Recall that $\ms_L$ corresponds to the chemical potential at 
which ``equal-area'' in $\pr$ of the {\it finite} system is observed. 
Shown in \fig{fss_mhu}(a) is $\ms_L$ versus $L$. Again, Borgs and Kotecky 
predict an exponential $L$-dependence given by \eq{eq:sc}. We may therefore 
estimate $\ms_\infty$ as before, by plotting $\ln R$ versus $L$, using for 
$\ms_\infty$ that value at which the data become linear. The result is shown in 
\fig{fss_mhu}(b). Again, the data follow the straight line quite well, and we 
conclude $\ms_\infty \approx 5.6536$.

It is also of interest to consider the distribution $P_L(S)$ at coexistence, 
with $S$ the nematic order parameter. Here, we define $S$ as the maximum 
eigenvalue of the orientational tensor
\begin{equation}\label{eq:qq}
  Q_{\alpha\beta} = \frac{1}{A} \sum_{i=1}^N \left( 2 d_{i\alpha} 
  d_{i\beta} - \delta_{\alpha\beta} \right), 
\end{equation}
with $d_{i\alpha}$ the $\alpha$ component ($\alpha = x,y$) of the orientation 
$\vom_i$ of molecule $i$, $\delta_{\alpha\beta}$ the Kronecker delta, $N$ the 
number of particles, and system area $A=L^2$. For an isotropic phase, in the 
thermodynamic limit, $S$ equals zero. For a ``true'' nematic phase, i.e.~with 
long-range order, one has $S>0$ in the thermodynamic limit. Note that we have 
normalized \eq{eq:qq} with the system area $A$, and not with $N$ as is usually 
done. The reason to normalize with respect to $A$ is that, in the GC ensemble, 
$N$ is a fluctuating quantity. Note also that, for a perfectly aligned phase, 
$S$ becomes identical to the particle density $N/A$.

Several distributions $P_L(S)$ are shown in \fig{pn}(b). We emphasize that all 
these distributions were obtained using that value of $\mu$ at which 
``equal-area'' in $\pr$ was obtained. The data of \fig{pn}(b) are quite 
stunning. On the one hand, we observe a pronounced shift of the lower-peak 
toward $S \to 0$. This is to be expected, since this peak corresponds to the 
isotropic phase. Note that the $L$-dependence of the isotropic peak does not 
reveal any physical information: an ideal gas of rods would reveal the same 
effect. The $L$-dependence of $S$ in the isotropic phase is purely a numerical 
artifact. It stems from the fact that $S$ is positive, since always the {\it 
maximum} eigenvalue of the orientational tensor is taken. Consequently, the 
distribution of $S$ in the isotropic phase {\it cannot} be gaussian around 
$S=0$. Instead, the distribution is ``skewed'', with the peak {\it always} being 
located at $S>0$. As the system size is increased, the isotropic peak shifts to 
zero, see the discussion by Eppenga and Frenkel \cite{eppenga.frenkel:1984}, 
precisely what we observe.

\begin{figure}
\begin{center}
\includegraphics[clip=,width=\columnwidth]{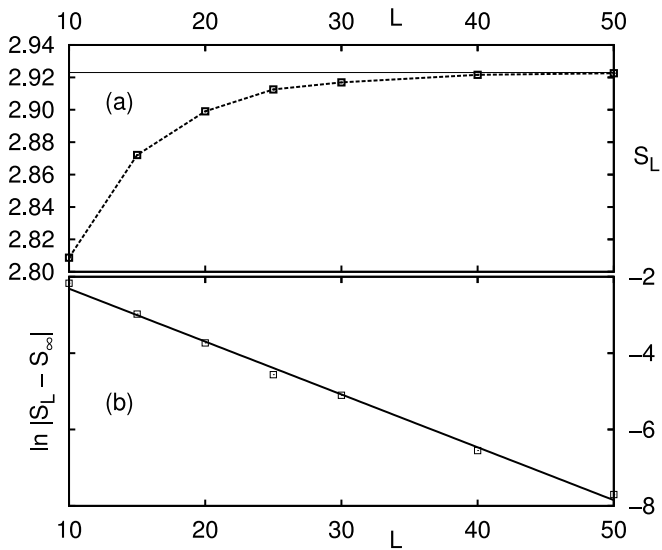}
\caption{\label{nem} Finite-size scaling of the position of the nematic peak in 
$P_L(S)$ of \fig{pn}(b). (a) Peak position $S_L$ as a function of $L$. The 
horizontal line marks the nematic order parameter $S_\infty$ in the 
thermodynamic limit. (b) Finite-size scaling analysis using \eq{eq:sc}. The 
straight line is a linear fit; see details in text.}
\end{center}
\end{figure} 

In contrast, on the scale of \fig{pn}(b), the nematic peak appears to be rather 
insensitive to the system size $L$. In fact, finite-size scaling of the nematic 
peak position suggests that a finite value $S_\infty$ in the thermodynamic limit 
is maintained, see \fig{nem}. Shown in \fig{nem}(a) is the nematic peak position 
$S_L$ as a function of $L$. The data are quite interesting, because they show an 
{\it increase} of nematic order with increasing system size $L$. Shown in 
\fig{nem}(b) is the result of the finite-size scaling analysis using \eq{eq:sc}. 
Again, $S_\infty$ was obtained by tuning, until the best collapse of the data 
onto a straight line occurred. For the nematic order parameter in the 
thermodynamic limit, we obtain $S_\infty \approx 2.923$. Note that, for $\nu=0$, 
the liquid crystal potential of \eq{eq:ol} is separable. For this special case, 
Straley has proved the {\it absence} of long-range nematic order in the 
thermodynamic limit \cite{physreva.4.675}. In other words, $S_\infty$ should 
become zero, while our finite-size analysis, in contrast, suggests that 
$S_\infty$ remains finite. For the XY-model in two dimensions, it has been shown 
that the decay of magnetic order with system size is so slow, one would need a 
sample ``the size of Texas'' \cite{PhysRevB.49.8811} to see it. The most likely 
explanation is therefore that something similar also takes place in our liquid 
crystal model, and that the data of \fig{nem}(a) will eventually ``turn-over'' 
and decay to zero.

\begin{figure}
\begin{center}
\includegraphics[clip=,width=\columnwidth]{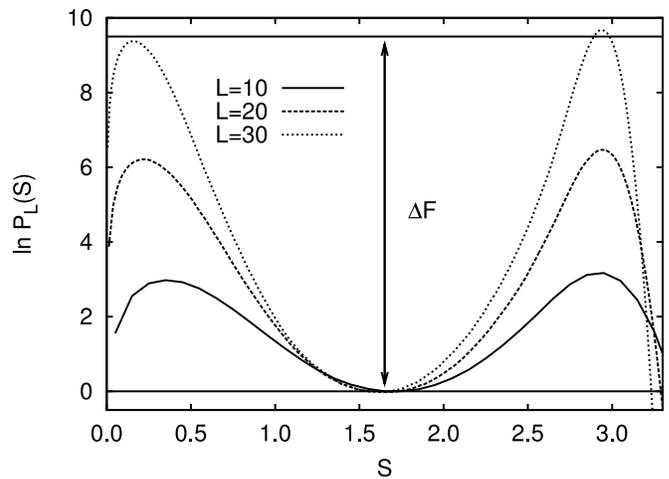}
\caption{\label{logps} Distributions $\ln P_L(S)$ for various system sizes 
$L$ as indicated. Also marked is $\Delta F$ for the $L=30$ system.}
\end{center}
\end{figure}

\begin{figure}
\begin{center}
\includegraphics[clip=,width=\columnwidth]{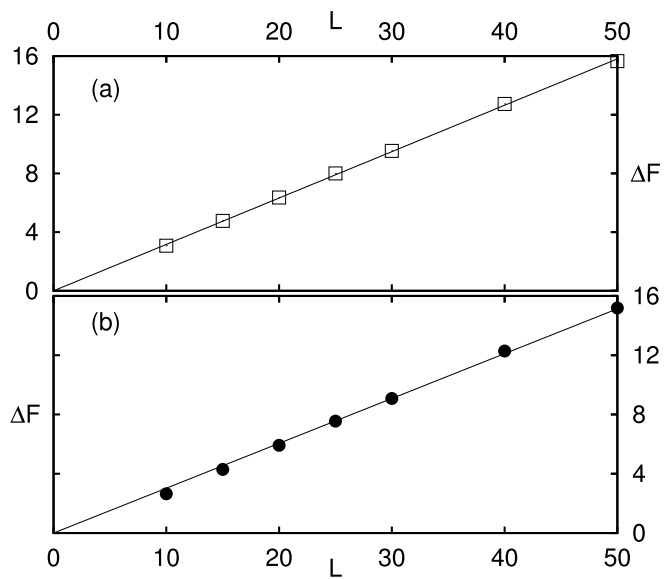}

\caption{\label{df} (a) Free energy barrier $\Delta F$, extracted from the 
distributions $\ln P_L(S)$ of \fig{logps}, as a function of $L$. The straight 
line is a fit to \eq{eq:lt}. (b) Same as (a), but with $\Delta F$ extracted from 
$\ln \pr$.}

\end{center}
\end{figure}

\begin{figure*}
\begin{center}
\includegraphics[clip=,width=18cm]{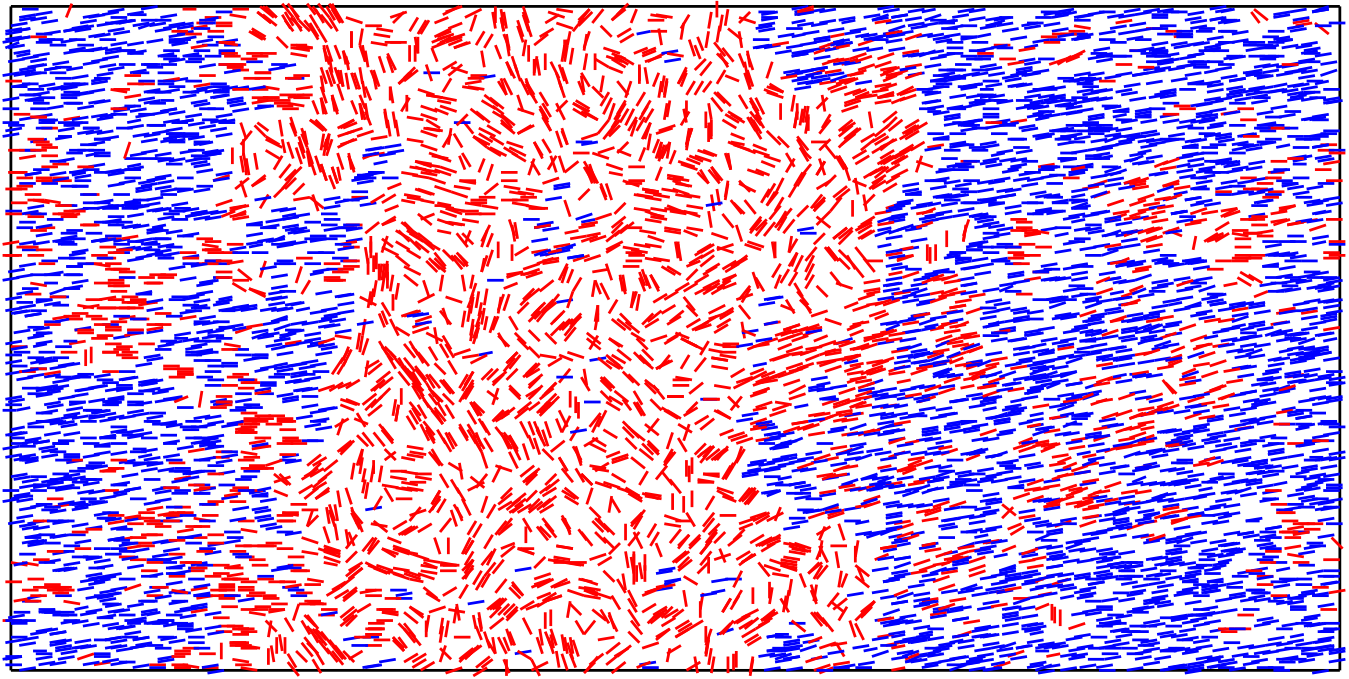}
\caption{\label{ss} Simulation snapshot obtained in the coexistence region for 
$p=60$. Each line segment represents a particle. Clearly visible is that the 
system has phase-separated into an isotropic domain, and a nematic domain (where 
we leave open the question whether the nematic phase exhibits true long-range 
order, or only quasi long-range order). Note also that the interfaces are not 
flat, and that they appear to be decorated with capillary waves.}
\end{center}
\end{figure*}

Next, we consider the logarithm of the distributions, which essentially reflect 
{\it minus} the free energy of the system. Shown in \fig{logps} is $\ln P_L(S)$ 
for several system sizes $L$. Again, we emphasize that the distributions were 
obtained using that value of $\mu$ at which ``equal-area'' in $\pr$ was 
obtained. For each distribution $\ln P_L(S)$, we may read-off the average peak 
height $\Delta F$, measured with respect to the minimum between the peaks. By 
increasing the size of the system, $\Delta F$ increases as well. As was shown by 
Binder, $\Delta F$ corresponds to the free-energy cost of having two interfaces 
in the system \cite{binder:1982}. More precisely, in two dimensions, we expect 
that
\begin{equation}\label{eq:lt}
  \Delta F = 2 \gamma L, 
\end{equation}
with $\gamma$ the line tension (the factor of two stems from the use of periodic 
boundary conditions, which lead to the formation of two interfaces in the 
system, see also the snapshot of \fig{ss}). Shown in \fig{df}(a) is $\Delta F$ 
as a function of $L$; the data are indeed well described by a straight line 
through the origin. From the slope of the line, we obtain $\gamma = 0.158 \, \kb 
T/a$. Of course, we may also read-off the barrier in $\ln \pr$, shown in 
\fig{df}(b). Again, a linear increase of $\Delta F$ is observed, and from the 
slope of the line we obtain $\gamma = 0.151 \, \kb T/a$, which is very close to 
our previous estimate.

In summary, and in agreement with our theoretical results, we find that an 
exponent $p=75$ is high enough to induce a first-order transition in the {\it 
off-lattice} liquid crystal model of \eq{eq:ol}. The scaling of the coexisting 
densities with system size are well described by what is expected for such a 
transition. The same also holds for the growth of the free energy barrier 
$\Delta F$. An interesting and unexpected result, which certainly requires 
further elaboration, is the scaling of the nematic order parameter, see 
\fig{nem}. If the trend continues for $L \to \infty$, long-range nematic order 
in a two-dimensional liquid crystal would, after all, be possible.

\subsection{Results for different $p$}

We have also measured the coexistence densities, chemical potential, and line 
tension for different values of $p$, while keeping $\nu=0$ and $\epsilon=2.5$. 
Here, we did not perform a detailed finite-size scaling analysis. Instead, the 
data for $p \neq 75$ were obtained in a single simulation of a (reasonably 
large) system. In \fig{dft-gaussbino}, we show the density of the isotropic 
phase, and of the nematic phase, as a function of $p$. Note that the region in 
between the curves corresponds to phase coexistence. If one performs a 
simulation in this region, for example by keeping the density fixed, snapshots 
strikingly reveal the two-phase coexistence, see \fig{ss}. Shown in 
\fig{dft-gaussmu} is the coexistence chemical potential $\ms$ versus $p$, 
compared to the DFT result. Finally, in \fig{dft-gvsp}(b), we show the line 
tension versus $p$.

\section{Discussion and Summary}

In summary, we have provided strong evidence that {\it off-lattice} liquid 
crystals in two dimensions can also undergo first-order phase transitions. To 
this end, the pair potential of \eq{eq:ol} was introduced, constructed to be 
purely repulsive, short-ranged, and to obey inversion symmetry. The first-order 
transition takes place when the pair potential becomes sufficiently ``sharp and 
narrow'', i.e.~for a sufficiently large exponent $p$ in \eq{eq:ol}. A simple DFT 
calculation puts the threshold value at around $p=8$. When $p>8$, the theory 
predicts a first-order transition from a low-density (isotropic) phase, to a 
high-density (nematic) phase. In other words, there is a finite density gap 
between the two phases, as well as a jump in the nematic order parameter.

An important conclusion of this work is that many of the trends predicted by the 
theory, also appear in the computer simulations. In other words, key properties 
of \eq{eq:ol} are already captured at the mean-field level. This finding is 
somewhat surprising because, in low spatial dimension, mean-field is typically 
assumed to be unreliable. The simulations, in agreement with the theory, find 
strong evidence of a first-order transition, already at $p>50$. Note that we 
have backed our simulations with a detailed investigation of finite-size 
effects, and that these were shown to be consistent with a first-order phase 
transition. We have not determined in our simulations the precise value of $p$ 
where the crossover from continuous to first-order occurs, since such 
simulations would be extremely time consuming, but we expect this will exceed 
the theoretical bound $p=8$ significantly.

Nevertheless, for those values of $p$ where the simulations do observe the 
first-order transition, a profound density gap between the two coexisting phases 
is found, see the phase diagram of \fig{dft-gaussbino}. In the limit of large 
$p$, theory and simulation are in qualitative agreement: both show an increase 
in density of the nematic phase with $p$, while the density of the isotropic 
phase is much less sensitive to $p$. At lower $p$, qualitative discrepancies 
arise, but these can be attributed to the large-$p$ approximation used by the 
theory, which obviously breaks down here. Interestingly, when comparing the 
coexistence chemical potential $\ms$ between theory and simulation, see 
\fig{dft-gaussmu}, the agreement is remarkably good, also at low $p$. 
Apparently, the breaking-down of the large-$p$ approximation does not affect 
$\ms$ as much as it does the coexistence densities. With regards to the line 
tension, see \fig{dft-gvsp}, the agreement between theory and simulation is 
merely qualitative: by increasing $p$, the tension increases in both cases, but 
the actual numerical values differ profoundly. A possible explanation may be the 
presence of capillary-wave interface fluctuations. These fluctuations are 
neglected in the theory, while the simulation snapshot of \fig{ss} suggests that 
interface fluctuations are actually quite strong.

A rather controversial result of this work is the value of the nematic order 
parameter $S$ in the nematic phase. The DFT result, see \fig{dft-s}, predicts 
rather large values of $S$, once the transition has become first-order. Of 
course, in realistic systems, one always has defects, which are expected to 
destroy nematic order in the thermodynamic limit. Since such defects are 
discarded by the theory, we expect that \fig{dft-s} is merely an artifact, and 
that realistic systems in the thermodynamic limit will always have $S=0$, 
regardless the value of $p$. Still, it is somewhat surprising that our computer 
simulations also suggest that $S>0$, see \fig{nem}, and that the finite-size 
effects in our data are {\it not} compatible with a decay of $S$ to zero. At 
this point, the most likely explanation is that the decay of $S$ with system 
size only shows-up in macroscopically large samples, which are clearly out of 
reach in any foreseeable simulation.

Note that our theory has also made a number of intriguing predictions for the 
case where the translational and orientational degrees of freedom are coupled, 
i.e.~when $\nu \neq 0$. In this case, strong anchoring effects are predicted, as 
well as the possibility of the isotropic-to-nematic transition being pre-empted 
by freezing. Since our particles are ultrasoft, the formation of stable 
aggregate or cluster mesophases as encountered in various soft-sphere systems 
\cite{mladek, glaser} is also possible. All these scenarios need to be verified 
by computer simulation. We are currently developing new simulation methodology 
to study anchoring effects at the isotropic-nematic interface; some preliminary 
information about the method is already available \cite{vink:2007}. The 
application of these new techniques to the 2D liquid crystal model of the 
present work is therefore postponed to a future publication.

Of course, it would be interesting if some of our findings could be confirmed in 
experiments. As we had already remarked in the Introduction, the condition of 
phase coexistence is obtained straightforwardly by keeping the density fixed at 
some value in the coexistence region. The problem will most likely be to achieve 
sufficiently ``sharp and narrow'' interactions. One possibility that we envision 
is to use a mixture of colloidal rods and non-adsorbing polymers. Despite the 
fact that colloidal rods cannot overlap, unlike the particles considered here, 
and that the addition of polymer renders the {\em effective} rod interactions 
{\em attractive}, which may induce an additional gas-liquid phase separation, we 
believe a first-order isotropic-nematic phase transition could be feasible. In 
particular, looking at the phase diagrams reported for 3D rod-polymer mixtures 
\cite{stroobants} we anticipate that the strong polymer-induced widening of the 
isotropic-nematic coexistence region carries over to 2D systems as well, and 
might induce a first-order transition. A prerequisite for this scenario is that 
both the size ratio (of rod length to polymer radius of gyration) and the 
polymer concentration are sufficiently large.

Finally, we would like to remind the reader that the original idea of this work, 
namely to use ``sharp and narrow'' interactions is not new, and goes back to the 
work of \olcite{physrevlett.52.1535}. Here, the approximate correspondence 
between generalized XY-models and the Potts model \cite{revmodphys.54.235} was 
already exploited to demonstrate the possibility of having a first-order 
transition in a 2D spin system with continuous degrees of freedom. The extension 
of this work has been to apply the same ideas to {\it off-lattice} liquid 
crystals. Nevertheless, the approximate link to the Potts model can still be 
uncovered, using $q \propto p^{1/2}$ \cite{physrevlett.88.047203}. Here, $p$ is 
the exponent in \eq{eq:ol}, and $q$ the number of Potts states. Note that this 
relation is valid only asymptotically for large $p$. Many of the results of this 
work, for example the cross-over from a continuous to a first-order transition 
(\fig{dft-df}), the variation of $\ms$ with $p$ (\fig{dft-gaussmu}), and even 
the increase of the line tension with $p$ (\fig{dft-gvsp}), have their analogues 
in the 2D Potts model (see for example \olcite{borgs.janke:1992}, where an 
explicit formula for the line tension of the Potts model is given).

\acknowledgments

This work was supported by the {\it Deutsche Forschungsgemeinschaft} under
the SFB-TR6 (project section D3). 

\bibstyle{revtex}
\bibliography{REFS_vink,REFS_rik}

\begin{thebibliography}{46}
\bibitem[{not({\natexlab{a}})}]{note1}
\bibinfo{note}{This can be shown by replacing $|\cos (\varphi^{\prime} -
  \varphi )|$ in \eq{coef} by the Onsager kernel $-|\sin (\varphi ^{\prime} -
  \varphi )|$, see \olcite{physreva.4.675}. Both kernels yield the same
  coefficient $c_2 = 4/(3 \pi)$, and hence the same bifurcation density
  $c^\star = 3 \pi/2$.}
\bibitem[{not({\natexlab{b}})}]{note2}
\bibinfo{note}{The $\vartheta$-dependence also vanishes for the bulk case, as
  it should. This can easily be verified by carrying out the total spatial
  integral of $\sigma$ [cf.~\eq{dft4}].}
\end{thebibliography}

\end{document}